# Simple arithmetic versus intuitive understanding:

# The case of the impact factor

Ronald Rousseau [1,2,3] and Loet Leydesdorff [4]

[1] KHBO (Association K.U.Leuven), Industrial Sciences and Technology, Zeedijk 101,
B-8400 Oostende, Belgium
E-mail: ronald.rousseau@khbo.be

[2] K.U.Leuven, Dept. Mathematics,
Celestijnenlaan 200B, B-3000 Leuven (Heverlee), Belgium

[3] Universiteit Antwerpen (UA), IBW, Stadscampus, Venusstraat 35,
B-2000 Antwerpen, Belgium

[4] Amsterdam School of Communication Research (ASCoR), University of Amsterdam,
Kloveniersburgwal 48, 1012 CX Amsterdam, The Netherlands
E-mail: loet@leydesdorff.net

**Abstract**

We show that as a consequence of basic properties of elementary arithmetic journal impact factors show a counterintuitive behaviour with respect to adding non-cited articles. Synchronous as well as diachronous journal impact factors are affected. Our findings provide a rationale for not taking uncitable publications into account in impact factor calculations, at least if these items are truly uncitable.

**Introduction**

In this note we show how simple arithmetic may influence our understanding of the impact factor. Concretely, it is possible that the impact factor of journal J is larger than the impact factor of journal J' and that adding the same number of non-cited articles to both reverses the mutual order. Although completely natural from a mathematical point of view, we consider such behaviour as counterintuitive. Indeed, journal J seems more visible than journal J': how can then adding non-cited items make journal J' more visible than journal J?

**Journal impact factors**

We recall the definitions of the synchronous and the diachronous journal impact factor. The n-year synchronous impact factor of journal J in year Y is defined as (Rousseau, 1988):

$$IF_n(J,Y) = \frac{\sum_{i=1}^{n} Cit(Y,Y-i)}{\sum_{i=1}^{n} Pub(Y-i)} = \frac{\frac{1}{n}\sum_{i=1}^{n} Cit(Y,Y-i)}{\frac{1}{n}\sum_{i=1}^{n} Pub(Y-i)} \qquad (1)$$

In this formula the number of citations received by journal J (from all members of the pool of sources under consideration) in the year Y, by articles published in the year X, is denoted as $CIT_J(Y,X)$, where, for simplicity we have not included the index J. Similarly, PUB(Z) denotes the number of articles published by this same journal in the year Z. We made it clear in equation (1) that the standard synchronous journal impact factor is a ratio of averages (RoA). Hence we will denote it as RAIF. When n = 2 one obtains the classical Garfield (1972) journal impact factor. Since a few years also the 5-year journal impact factor is provided in Thomson Reuters' Web of Science. The term 'synchronous' refers to the fact that the citation data used to calculate it are data collected in the same year. We next recall the definition of the diachronous impact factor.

The n-year diachronous impact factor of journal J for the year Y, denoted as $IMP_n(J,Y)$, is defined as

$$IMP_n(J,Y) = \frac{\sum_{i=s}^{s+n-1} Cit(Y+i,Y)}{Pub(Y)} \qquad (2)$$

where *s* = 0 or 1, depending on whether one includes the year of publication or not. The term 'diachronous' refers to the fact that the data that are used to calculate this impact factor derive from a number of different years with a starting point somewhere in the past and encompassing subsequent years (Ingwersen *et al.*, 2001).

**Ranking invariance with respect to non-cited items**

We consider the following form of invariance. If a performance indicator I is calculated for journals J and J' and I(J) < I(J') then, if we add the same number of publications with zero citations, we require that also for the new situation I(J) < I(J'). We refer to this requirement as ranking invariance with respect to non-cited items. This notion is totally different from the consistency notions introduced by Waltman and van Eck (Waltman & van Eck, 2009; Waltman et al., 2011) or by Marchant (2009) (under the name of independence). Recall that, for good reasons, the notion of consistency as defined by these authors refers to cases where the number of publications (in the denominator) is the same for both journals. We do not require this.

Next we show that impact factors are not ranking invariant with respect to non-cited items. Consider the following example (see Table 1).

Table 1: Data for the calculation of the Garfield impact factor (RoA case) for the year Y.

|           | J        | J'       |
|-----------|----------|----------|
| Pub(Y-1)  | 10 (+25) | 30 (+25) |
| Pub(Y-2)  | 10       | 30       |
| Cit(Y,Y-1)| 30       | 60       |
| Cit(Y,Y-2)| 30       | 60       |

On the basis of Table 1, the Garfield impact factors of journals J and J' are $IF_2(J,Y) = 3$ and $IF_2(J',Y) = 2$, so that $IF_2(J,Y) > IF_2(J',Y)$. However, adding 25 non-cited publications, yields the new impact factors: $IF_2(J,Y) = 60/45 = 1.33$ and $IF_2(J',Y) = 120/85 = 1.41$, so that for the new situation the relation between the impact factors reverses.

We note that for the classical synchronous impact factor $IF_2$ ranking invariance with respect to non-cited items always holds for the special case that both $IF(J_1) < IF(J_2)$ and the sum of Cit(Y,Y-1) and Cit(Y,Y-2) is smaller (or equal) for journal $J_1$ than for journal $J_2$. Indeed: denoting the 2-year impact factor of journal $J_i$ (i = 1,2) simply by $C_i/P_i$ we have:

$$\frac{C_1}{P_1} < \frac{C_2}{P_2} \text{ and } C_1 \leq C_2$$

If now Z denotes the added number of publications with no citations we have to show that:

$$\frac{C_1}{P_1 + Z} < \frac{C_2}{P_2 + Z} \Leftrightarrow C_1(P_2 + Z) < C_2(P_1 + Z) \Leftrightarrow C_1 P_2 + C_1 Z < C_2 P_1 + C_2 Z$$

This is clearly true since $C_1 P_2 < C_2 P_1$ and $C_1 Z \leq C_2 Z$.

We also note that if $\frac{C_1}{P_1} < \frac{C_2}{P_2}$ and $P_1 \leq P_2$ then $\frac{C_1}{C_2} < \frac{P_1}{P_2} \leq 1$, hence also $C_1 < C_2$. This implies that also under these conditions ranking invariance with respect to non-cited items holds. Denoting the impact factor of journal J when Z non-cited items are added by $IF_Z(J)$ brings us to the following result.

Proposition. If $IF(J_1) < IF(J_2)$ then rank reversal, i.e. $IF_Z(J_1) > IF_Z(J_2)$ occurs if and only if $C_1 > C_2$ and $Z > \frac{D}{C_1-C_2}$, where $D = P_1C_2 - P_2C_1$.

Proof. We already know that if $C_1 \leq C_2$ then there is no rank reversal. If now $IF(J_1) < IF(J_2)$ this implies that $P_1C_2 > P_2C_1$. Its (positive) difference $P_2C_1 - P_1C_2$ is denoted as D. Now $IF_Z(J_1) > IF_Z(J_2) \Leftrightarrow (P_1+Z)C_2 < (P_2+Z)C_1 \Leftrightarrow P_2C_1 + D + ZC_2 < P_2C_1 + ZC_1 \Leftrightarrow D < (C_2-C_1)Z\ ) \Leftrightarrow Z > \frac{D}{C_1-C_2}$.

A simple variation of Table 1 shows that also the diachronous impact does not satisfy this property either, see Table 2.

Table 2: Data for the calculation of the dynamic impact factor (RoA case) for the year Y.

|            | J        | J'       |
|------------|----------|----------|
| Pub(Y)     | 20 (+25) | 60 (+25) |
| Cit(Y,Y)   | 10       | 20       |
| Cit(Y,Y+1) | 20       | 40       |
| Cit(Y,Y+2) | 30       | 60       |

With s = 0, we have $IMP_3(J,Y) = 60/20 = 3$ and $IMP_3(J',Y) = 120/60 = 2$. Adding 25 non-cited publications yields the new diachronous impact factors: $IMP_3(J,Y) = 60/45 = 1.33$ and $IMP_3(J',Y) = 120/85 = 1.41$. The rank-order of the two journals in terms of their impact factor thus is reversed by adding an equal number of non-cited items to both.

**AoR versus RoA**

We have shown that the standard synchronous impact factor is of the RoA-form and that it does not satisfy ranking invariance with respect to non-cited items. Let us analyze whether perhaps an Average of Ratios (AoR) form of the synchronous impact factor behaves better in this respect. First we define the ARIF as:

$$ARIF_n(J,Y) = \frac{1}{n}\sum_{i=1}^{n} \frac{Cit(Y,Y-i)}{Pub(Y-i)} \qquad (3)$$

However, it turns out that the AoR-form behaves even worse with respect to ranking invariance. Indeed, consider the case of a two-year impact factor ($ARIF_2$) and assume that journals J and J' have in each year equal numbers of publications. If $RAIF_2(J,Y) > RAIF_2(J',Y)$, (or, $IF_2(J,Y) > IF_2(J',Y)$), this means that journal J received more citations than journal J' (in the year Y). Adding the same number of zero-cited publications to both, does not change the total number of citations received, and hence J's standard impact factor remains smaller than J' (of course both impact factors decrease by increasing the denominators). The same argument holds for the n-year synchronous impact factor (RA-case). This, however, does not hold for ARIF. Consider the example shown in Table 3.

Table 3. Data for the calculation of a two-year synchronous impact factor (AoR case) for the year Y.

|          | J        | J'       |
|----------|----------|----------|
| Pub(Y-1) | 30 (+10) | 30 (+10) |
| Pub(Y-2) | 20       | 20       |
| Cit(Y,Y-1) | 10     | 120      |
| Cit(Y,Y-2) | 80     | 10       |

Based on the data shown in Table 3, we have: $ARIF_2(J,Y) = (0.5).(10/30+80/20) = 2.17$ and $ARIF_2(J',Y) = (0.5).(120/30+10/20) = 2.25$ so that $ARIF_2(J,Y) < ARIF_2(J',Y)$. However, adding 10 publications in the year Y-1 yields the new impact factors: $ARIF_2(J,Y) = (0.5).(10/40+80/20) = 2.13$ and $ARIF_2(J',Y) = (0.5).(120/40 + 10/20) = 1.75$, so that for the new situation $ARIF_2(J,Y) > ARIF_2(J',Y)$. ARIF is more sensitive to adding publications with no citations to the denominator than RAIF *because* ARIF is an average (cf. Ahlgren et al., 2003); RAIF, however, is not an average, but a quotient between two summations (Egghe & Rousseau, 1996).

It is easy to find similar examples of violations against the assumption of ranking invariance for any n-synchronous impact factor calculated in the AoR way.

**A remark concerning the framework of impact factor calculations: uncitable items**

When Garfield introduced the impact factor, he decided to introduce the notion of uncitable items. The idea was that journals should not be 'punished' for publishing obituaries, corrections, editorials and similar types of publications, which usually receive no or few citations. Although this seems reasonable, there are in practice two problems with this notion. One is to decide which publications are uncitable, and the other one is the fact that Garfield also decided to include citations to these 'uncitable' articles – when they occur - to the total number of received articles. It has been shown, see e.g. (Moed & van Leeuwen, 1995) that this practice may lead to serious distortions in journal impact.

Assume now that if uncitable items could be defined unambiguously, and that they are really never cited, which way of calculating an impact factor is then better? Taking all publications into account (including the – uncited – uncitable ones), or taking only the 'citable' ones (cited or not)? The answer is clearly that the second method should be used, as otherwise it would be possible that journal J obtains a higher impact than journal J' due to uncitable publications.

**Conclusion**

We have shown that, as a consequence of simple arithmetic, not satisfying the requirement of ranking invariance with respect to non-cited items, is a normal mathematical property related to taking ratios. For the calculation of synchronous impact factors, the standard RoA approach is to be preferred above the AoR approach, as the RoA approach satisfies ranking invariance with respect to non-cited items for journals with the same number of publications, while the AoR approach may fail even in this case. Our findings provide a rationale for not taking uncitable publications into account in impact factor calculations, at least if these items are truly uncitable, that is, are really never cited. Furthermore, they provide another argument against using averages in the case of highly skewed distributions (Ahlgren et al., 2003; Bornmann & Mutz, 2011; Leydesdorff & Opthof, 2011).

Acknowledgements

The authors thank Raf Guns (UA), Wolfgang Glänzel and two anonymous reviewers for helpful suggestions.

**References**

Ahlgren, P., Jarneving, B., & Rousseau, R. (2003). Requirement for a Cocitation Similarity Measure, with Special Reference to Pearson's Correlation Coefficient. *Journal of the American Society for Information Science and Technology, 54*(6), 550-560.

Bornmann, L., & Mutz, R. (2011). Further steps towards an ideal method of measuring citation performance: The avoidance of citation (ratio) averages in field-normalization. *Journal of Informetrics, 5*(1), 228-230.

Egghe, L., & Rousseau, R. (1996). Averaging and globalising quotients of informetric and scientometric data. *Journal of Information Science, 22*(3), 165.

Garfield, E. (1972). Citation analysis as a tool in journal evaluation. *Science 178* (Number 4060), 471-479.

Ingwersen, P., Larsen, B., Rousseau, R. & Russell, J. (2001). The publication-citation matrix and its derived quantities. *Chinese Science Bulletin*, 46(6), 524-528.

Leydesdorff, L., & Opthof, T. (2011). Remaining problems with the "New Crown Indicator" (MNCS) of the CWTS. *Journal of Informetrics, 5*(1), 224-225.

Marchant, T. (2009). An axiomatic characterization of the ranking based on the h-index and some other bibliometric rankings of authors. *Scientometrics, 80*(2), 325-342.

Moed, H.F. & van Leeuwen, Th. N. (1995). Improving the accuracy of institute for scientific information's journal impact factors. *Journal of the American Society for Information Science*, 46(6), 461-467.

Rousseau R. (1988). Citation distribution of pure mathematics journals. In: *Informetrics 87/88* (Egghe L. & Rousseau R., eds). Amsterdam: Elsevier, pp. 249-262.

Waltman, L. & van Eck, J.N. (2009). A taxonomy of bibliometric performance indicators based on the property of consistency. In: *ISSI 2009* (B. Larsen & J. Leta, eds.). Sao Paulo: BIREME & Federal University of Rio de Janeiro, pp. 1002-1003.

Waltman, L., Van Eck, N. J., Van Leeuwen, T. N., Visser, M. S., & Van Raan, A. F. J. (2011). Towards a new crown indicator: some theoretical considerations. *Journal of Informetrics, 5*(1), 37-47.